\documentclass[aps,twocolumn,prd,showpacs,showkeys,preprintnumbers,superscriptaddress,nobibnotes,floatfix,longbibliography]{revtex4-1}

\pdfoutput=1

\usepackage{amsmath}
\usepackage{amsfonts}
\usepackage{amssymb}
\usepackage{mathrsfs}
\usepackage{graphicx}
\usepackage{color}
\usepackage{longtable}
\usepackage{bm}
\usepackage{blindtext}
\usepackage{wasysym}
\usepackage{hyperref}
\usepackage[normalem]{ulem}

\newcommand{\ie}{{\it i.e.}}

\newcommand{\eg}{{\it e.g.}}

\newcommand{\eq}{Eq.}

\newcommand{\fig}{Fig.}

\newcommand{\Ref}{Ref.}

\newcommand{\equ}[1]{\eq~(\ref{equ:#1})}
\newcommand{\figu}[1]{\fig~\ref{fig:#1}}

\newcommand{\bi}{\begin{itemize}}
\newcommand{\ei}{\end{itemize}}

\newcommand{\matr}[1]{\mathbf{#1}} 

\begin{document}
 
\title{Universe's Worth of Electrons to Probe Long-Range Interactions\\ of High-Energy Astrophysical Neutrinos}

\author{Mauricio Bustamante}
\affiliation{Niels Bohr International Academy and Discovery Center, Niels Bohr Institute, Blegdamsvej 17, 2100 Copenhagen, Denmark}
\author{Sanjib Kumar Agarwalla}
\affiliation{Institute of Physics, Sachivalaya Marg, Sainik School Post, Bhubaneswar 751005, India}
\affiliation{Homi Bhabha National Institute, Anushakti Nagar, Mumbai 400085, India}
\affiliation{International Centre for Theoretical Physics, Strada Costiera 11, 34151 Trieste, Italy\\
{\tt mbustamante@nbi.ku.dk (0000-0001-6923-0865), sanjib@iopb.res.in (0000-0002-9714-8866)} \smallskip}

\preprint{IP/BBSR/2018-13}

\date{September 27, 2018}

\begin{abstract}
 Astrophysical searches for new long-range interactions complement collider searches for new short-range interactions.  Conveniently, neutrino flavor oscillations are keenly sensitive to the existence of long-ranged flavored interactions between neutrinos and electrons, motivated by lepton-number symmetries of the Standard Model.  For the first time, we probe them using TeV--PeV astrophysical neutrinos and accounting for all large electron repositories in the local and distant Universe.  The high energies and colossal number of electrons grant us unprecedented sensitivity to the new interaction, even if it is extraordinarily feeble.  Based on IceCube results for the flavor composition of astrophysical neutrinos, we set the ultimate bounds on long-range neutrino flavored interactions.
\end{abstract}


\maketitle


{\bf Introduction.---}  Are there fundamental interactions whose range is macroscopic but finite?  
New interactions with ranges of up to 1~A.U. are severely constrained\ \cite{Lee:1955vk, Okun:1995dn, Williams:1995nq, Dolgov:1999gk, Adelberger:2003zx, Williams:2004qba}: they are feeble at best, so testing for them is tough.  Still, searches for new long-range interactions vitally complement collider searches for new short-range interactions.

\begin{figure}[ht!]
 \centering
 \includegraphics[width=\columnwidth]{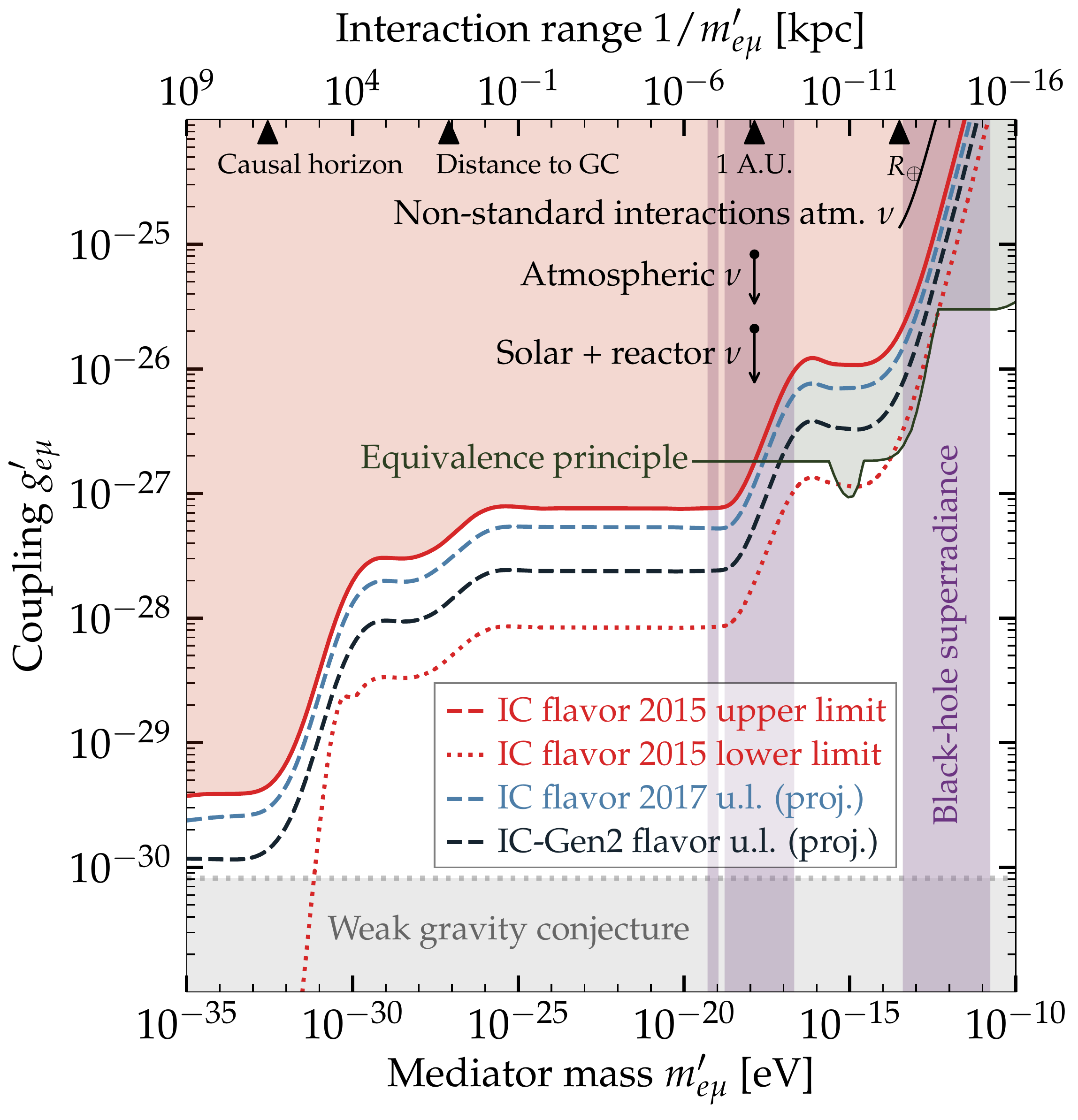}
 \caption{\label{fig:limits_emu}Constraints on the $Z_{e\mu}^\prime$ boson mediating long-range neutrino-electron interactions.  Our limits come from the flavor composition of high-energy astrophysical neutrinos at $1\sigma$, using current IceCube results and projections for IceCube and IceCube-Gen2, assuming normal neutrino mass ordering and a spectrum $\propto E_\nu^{-2.5}$.  Existing direct limits are from atmospheric\ \cite{Joshipura:2003jh}, and solar and reactor neutrinos\ \cite{Bandyopadhyay:2006uh}.  Indirect limits, from searches for non-standard neutrino interactions\ \cite{Mitsuka:2011ty, Ohlsson:2012kf, Gonzalez-Garcia:2013usa} (90\% C.L.), tests of the equivalence principle\ \cite{Schlamminger:2007ht} (95\% C.L.), and black-hole superradiance\ \cite{Baryakhtar:2017ngi} (90\% C.L.).  The weak gravity conjecture\ \cite{ArkaniHamed:2006dz} suggests that gravity is the weakest force and so $g_{e\mu}^{\prime 2} \geq G_N m_\nu^2$; we adopt a neutrino mass $m_\nu = 0.01$~eV.}
 \vspace*{-.55cm}
\end{figure}

We present a novel way to study long-range interactions between neutrinos and electrons.  Neutrinos are fitting test particles: in the Standard Model (SM), they interact only weakly, so the presence of a new interaction could more clearly stand out.  By considering interaction ranges up to cosmological scales, we become sensitive to the largest electron repositories in the local and distant Universe: the Earth, Moon, Sun, Milky Way, and cosmological electrons.  The collective effect of the colossal number of electrons grants us unprecedented sensitivity even if their individual contribution is feeble.

Symmetries of the SM naturally motivate considering new neutrino-electron interactions.  In the SM, lepton number $L_l$ ($l=e,\mu,\tau$) --- the number of leptons minus anti-leptons of flavor $l$ --- is conserved.  So are certain combinations of lepton numbers --- among them, $L_e-L_\mu$ and $L_e-L_\tau$.  Yet, when treated as broken local symmetries, they introduce a new interaction between electrons, $\nu_e$, and either $\nu_\mu$ or $\nu_\tau$,
mediated by a new neutral vector boson with undetermined mass and coupling\ \cite{He:1990pn,He:1991qd,Foot:1994vd}.  If the boson is light, the range of the interaction is long. 

The new interaction affects neutrino oscillations; at high energies, it might drive them.  Thus, for the first time, we look for signs of it in the TeV--PeV astrophysical neutrinos seen by IceCube\ \cite{Aartsen:2013bka,Aartsen:2013jdh,Aartsen:2013eka,Aartsen:2014gkd,Aartsen:2014muf,Aartsen:2015xup,Aartsen:2015knd,Aartsen:2015rwa,Aartsen:2016xlq}, whose flavor composition is set by oscillations that occur en route to Earth.

Figure \ref{fig:limits_emu} shows that our limits on the new coupling are the strongest for mediator masses under $10^{-18}$~eV --- or interaction ranges above 1~A.U.  By exploring the parameter space continuously, down to masses of $10^{-35}$~eV, we improve by orders of magnitude over the reach of previous limits from atmospheric, solar, and reactor neutrino experiments\ \cite{Joshipura:2003jh, GonzalezGarcia:2006vp, Bandyopadhyay:2006uh, Samanta:2010zh, Davoudiasl:2011sz, Chatterjee:2015gta, Wise:2018rnb, Khatun:2018lzs}.  By tapping into a Universe's worth of electrons, we reach the best possible sensitivity.


{\bf Lepton-number symmetries.---}  We focus on the lepton-number symmetries $L_e-L_\mu$ and $L_e-L_\tau$ of the SM. The related symmetry $L_\mu-L_\tau$ --- which we do not consider here --- has been studied extensively as a means to generate a lepton mixing angle $\theta_{23} \approx 45^\circ$\ \cite{Choubey:2004hn, Ota:2006xr, Heeck:2011wj, Harigaya:2013twa, Altmannshofer:2014cfa, Heeck:2014qea, Crivellin:2015mga}.  These are anomaly-free symmetries\ \cite{He:1990pn,He:1991qd,Foot:1994vd}: when promoted to local $U(1)$ symmetries and broken, they produce some of the simplest extensions of the SM.  They only increase the particle content by adding one new neutral vector gauge boson, $Z_{e\mu}^\prime$ or $Z_{e\tau}^\prime$.  These acquire a mass $m_{e \beta}^\prime = g_{e \beta}^\prime \langle S_{e \beta} \rangle$ ($\beta = \mu, \tau$) by coupling to a scalar Higgs field with vacuum expectation value $\langle S_{e \beta} \rangle$\ \cite{He:1991qd,Foot:1994vd}.  In this prescription, $L_e-L_\beta$ remain global symmetries, and the undetermined values of $m_{e\beta}^\prime$ and $g_{e\beta}^\prime$ can be arbitrarily small.


\begin{figure}[t!]
 \centering
 \includegraphics[width=0.9\columnwidth]{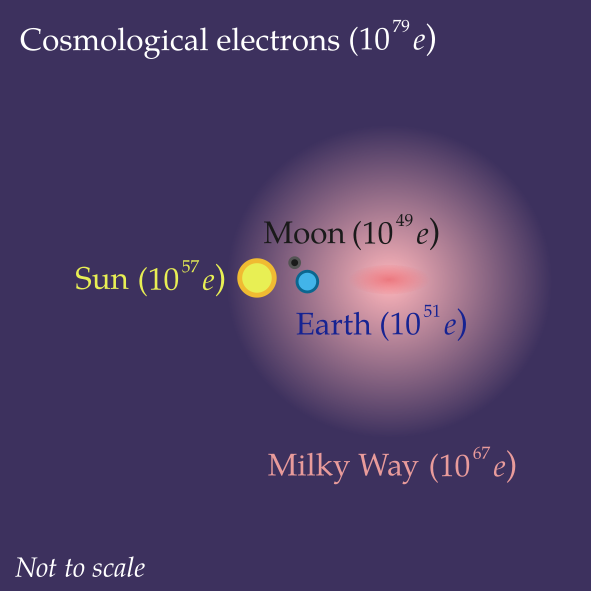}
 \caption{\label{fig:electron_sources}Electron repositories in the local and distant Universe used to set limits on long-range neutrino-electron interactions.}
\end{figure}

\begin{figure}[t!]
 \centering
 \includegraphics[width=\columnwidth]{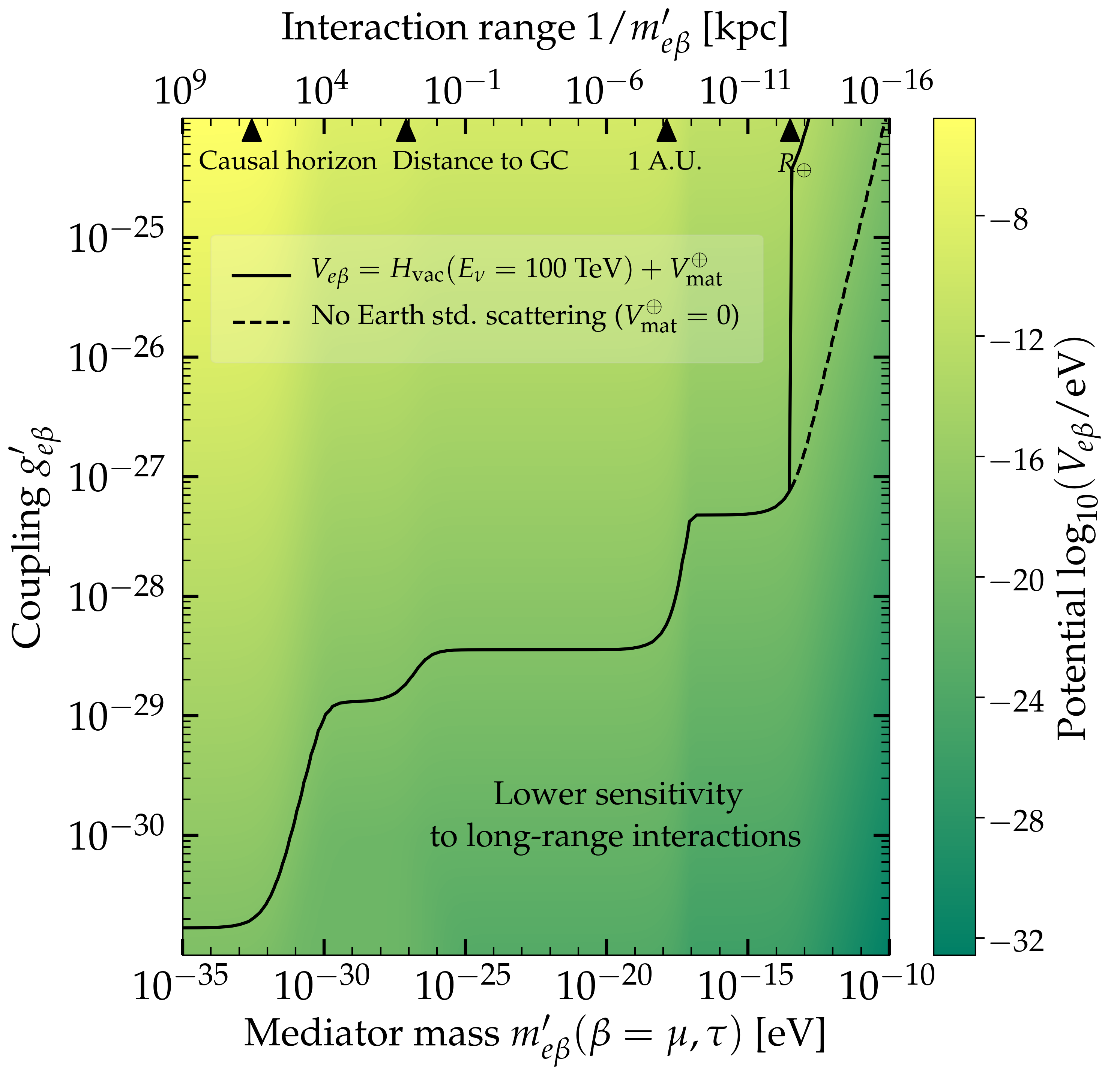}
 \caption{\label{fig:potential}Long-range potential $V_{e\beta}$ induced by the $L_e-L_\beta$ symmetry ($\beta = \mu, \tau$), sourced by electrons in the Earth, Moon, Sun, Milky Way, and by cosmological electrons.  The $Z_{e\beta}^\prime$ boson that mediates the potential has mass $m_{e\beta}^\prime$ and coupling $g_{e\beta}^\prime$.  The curve is the iso-contour of the potential at a value the vacuum oscillation Hamiltonian --- concretely, its element $H_{{\rm vac},ee}$ --- evaluated at $E_\nu = 100$~TeV, plus the potential $V_{\rm mat}^\oplus$ due to standard matter effects inside the Earth.  Due to the $\sim$$1/E_\nu$ dependence of $\matr{H}_{\rm vac}$ and the $\sim$$g_{e\beta}^{\prime 2}$ dependence of $V_{e\beta}$, the iso-contour would shift to lower couplings at higher $E_\nu$.}
\end{figure}

{\bf Long-range potential.---}  Under the $L_e-L_\beta$ symmetry, a neutrino separated a distance $d$ from a source of $N_e$ electrons experiences a Yukawa potential $V_{e\beta} = g_{e\beta}^{\prime 2} N_e (4 \pi d)^{-1} e^{-d / m_{e\beta}^\prime}$, mediated by the $Z_{e\beta}^\prime$.  The suppression due to the mediator mass kicks in at distances beyond the interaction range $1/m_{e\beta}^\prime$.  Thus, for a given value of the mass, the total potential is the aggregated contribution from all electrons located roughly within the interaction range.  We explore masses from $10^{-10}$~eV to $10^{-35}$~eV; the associated interaction range varies from meters to $10^3$~Gpc --- much larger than the observable Universe, \ie, effectively infinite.  Below, we outline the calculation of the potential; details are in the Supp.\ Mat.

Figure \ref{fig:electron_sources} sketches the electron repositories used in our analysis.  In the local Universe, the largest repositories of electrons are the Earth ($N_{e,\oplus} \sim 10^{51}$), Moon ($N_{e,\leftmoon} \sim 10^{49}$), Sun ($N_{e,\astrosun} \sim 10^{57}$), and the stars and gas of the Milky Way ($N_{e,{\rm MW}} \sim 10^{67}$).  For the Earth, we calculate the potential due to electrons in its interior acting on neutrinos that reach the detector from all directions, each traversing a different electron column density inside the Earth.  For the Moon and the Sun, we take them as point sources of electrons at distances of $d_{\leftmoon} \approx 4 \cdot 10^5$~km and $d_{\astrosun} = 1$~A.U.  For the Milky Way, we compute the potential at the position of the Earth --- 8~kpc from the Galactic Center (GC) --- due to all known Galactic baryonic matter.  We adopt a sophisticated model of the Galaxy that includes the central bulge, thin disc, and thick disc of stars and cold gas\ \cite{McMillan:2011wd}, and the diffuse halo of hot gas\ \cite{Miller:2013nza}. 

In addition, there is a cosmological contribution, previously overlooked, from $N_{e,{\rm cos}} \sim 10^{79}$ electrons contained inside the causal horizon\ \cite{Weinberg:2008zzc}, \ie, the largest causally connected region centered on the neutrino.  We gain sensitivity to these electrons when the interaction range is of Gpc-scale or larger.  Since the number density of cosmological electrons changes as the Universe expands, we compute a redshift-averaged potential due to them, weighed by the number density $\rho_{\rm src}$ of neutrino sources: $\langle V_{e\beta}^{\rm cos} \rangle \propto \int dz~ \rho_{\rm src}(z) \cdot dV_{\rm c}/dz \cdot V_{e\beta}^{\rm cos}(z)$, where $V_{e\beta}^{\rm cos}(z)$ is the potential at redshift $z$ and $V_{\rm c}$ is the comoving volume\ \cite{Hogg:1999ad}.  Because astrophysical neutrinos are largely extragalactic in origin\ \cite{Aartsen:2017ujz}, we reasonably assume that $\rho_{\rm src}$ follows the star formation rate\ \cite{Hopkins:2006bw, Yuksel:2008cu, Kistler:2009mv}. 

Figure \ref{fig:potential} shows the total potential $V_{e\beta} = V_{e\beta}^\oplus + V_{e\beta}^{\leftmoon} + V_{e\beta}^{\astrosun} + V_{e\beta}^{\rm MW} + \langle V_{e\beta}^{\rm cos} \rangle$ as a function of the mediator mass and coupling.  Tracing the iso-contour of constant $V_{e\beta}$ from high to low masses reveals the transitions that the potential undergoes as the interaction range grows.  From $10^{-10}$~eV to $10^{-18}$~eV, the potential is sourced mainly by the Earth and, to a lesser degree, the Moon.  The sharp jump at $1/m_{e\beta}^\prime = R_\oplus$ is due to standard Earth matter effects turning on.  At $10^{-18}$~eV, the interaction range reaches the Sun, the potential receives the contribution of solar electrons, and the iso-contour jumps to a lower value of the coupling.  At progressively smaller masses, the interaction range grows and the potential receives the aggregated contribution from electrons distributed in the Milky Way.  At $10^{-27}$~eV, the interaction range reaches the GC and the iso-contour jumps to an even lower value of the coupling, since the GC contains more electrons.  Finally, at $5 \cdot 10^{-33}$~eV, the interaction range reaches the size of the causal horizon, and the potential is saturated by all of the electrons in the observable Universe.


{\bf Flavor transitions.---}  The new interaction affects the evolution of flavor as neutrinos propagate.  The evolution is described by the Hamiltonian $\matr{H}_{e\beta} = \matr{H}_{\rm vac} + \matr{V}_{e\beta} + \Theta(R_\oplus-m_{e\beta}^{\prime -1})  \matr{V}_{\rm mat}^\oplus$, here written in the flavor basis.  The first term accounts for vacuum oscillations: $\matr{H}_{\rm vac} = (2E_\nu)^{-1} \matr{U} \matr{M}^2 \matr{U}^\dagger$, where $E_\nu$ is the neutrino energy, $\matr{M}^2 = \text{diag}(0, \Delta m_{21}^2, \Delta m_{31}^2)$, and $\matr{U}$ is the Pontecorvo-Maki-Nakagawa-Sakata (PMNS) mixing matrix, parametrized, as usual, via the mixing angles $\theta_{12}$, $\theta_{23}$, $\theta_{13}$, and the CP-violation phase $\delta_{\rm CP}$.  The second term accounts for the new interaction\ \cite{Joshipura:2003jh, GonzalezGarcia:2006vp, Bandyopadhyay:2006uh, Samanta:2010zh, Chatterjee:2015gta, Khatun:2018lzs, Wise:2018rnb}: $\mathbf{V}_{e\beta} = \text{diag} \left( V_{e\beta}, - \delta_{\mu\beta} V_{e\beta}, - \delta_{\tau\beta} V_{e\beta} \right)$.  The third term accounts for standard matter effects inside the Earth: $\matr{V}_{\rm mat}^\oplus = \text{diag}(V_{\rm mat}^\oplus, 0, 0)$, where $V_{\rm mat}^\oplus \equiv \sqrt{2} G_{\rm F} n_e^\oplus$ and $n_e^\oplus$ is the electron number density; see the Supp.\ Mat.\ for details.  This term is relevant only when the interaction range is smaller than the radius of the Earth, \ie, when $m_{e\beta}^{\prime -1} \leq R_\oplus$.  When the new potential or the standard matter potential dominates, the Hamiltonian becomes diagonal and flavor mixing turns off.  For anti-neutrinos, $\delta_{\rm CP} \to -\delta_{\rm CP}$, $\matr{V}_{e\beta} \to - \matr{V}_{e\beta}$, and $\matr{V}_{\rm mat}^\oplus \to - \matr{V}_{\rm mat}^\oplus$.

From here, we compute the probability of the flavor transition $\nu_\alpha \to \nu_\beta$.  For high-energy neutrinos, the probability oscillates rapidly with distance --- the oscillation length is tiny compared to the propagated distances, \ie, $10^{-10}$~Mpc vs.\ Gpc.  Thus, we approximate the probability by its average value\ \cite{Pakvasa:2008nx}, $P_{\alpha\beta}(E_\nu) = \sum_{i=1}^3 \lvert U^\prime_{\alpha i}(E_\nu) \rvert^2 \lvert U^\prime_{\beta i}(E_\nu) \rvert^2$, where $\matr{U}^\prime$ is the matrix that diagonalizes $\matr{H}_{e\beta}$.  It has the same structure as the PMNS  matrix, but its elements depend not only on $\theta_{12}$, $\theta_{23}$, $\theta_{13}$, and $\delta_{\rm CP}$, but also on $\Delta m_{21}^2$, $\Delta m_{31}^2$, $g_{e\beta}^\prime$, $m_{e\beta}^\prime$, and $E_\nu$.  Below, to obtain our results, we numerically compute $P_{\alpha\beta}$ for each choice of values of these parameters.

\begin{figure}[t!]
 \centering
 \includegraphics[width=\columnwidth]{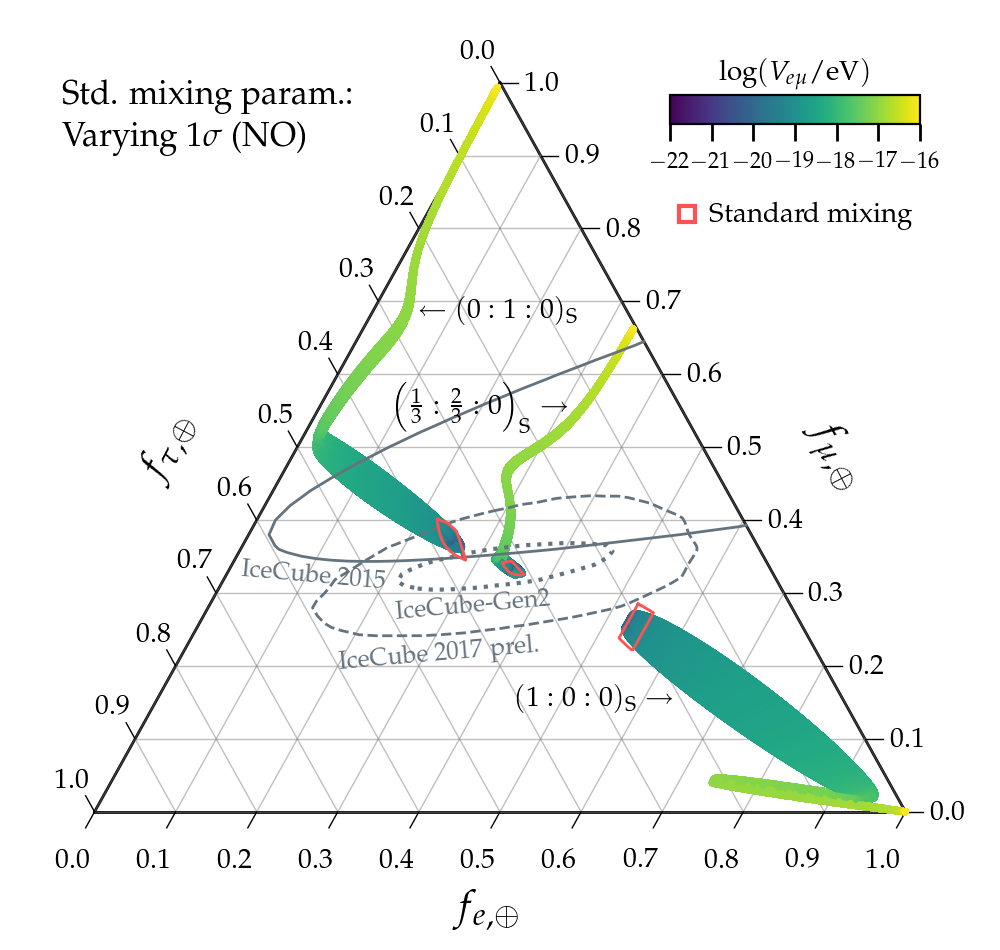}
 \caption{\label{fig:f_E_vary_V}Flavor ratios at Earth $f_{\alpha,\oplus}$ as functions of the long-range potential $V_{e\mu}$ associated to the $L_e-L_\mu$ symmetry, for three illustrative choices of flavor ratios at the sources $(f_{e, {\rm S}}:f_{\mu, {\rm S}}:f_{\tau, {\rm S}}) = \left( \frac{1}{3}:\frac{2}{3}:0 \right)$ (nominal case), $(0:1:0)$ (shown in Supp.\ Mat.) and $(1:0:0)$ (pure-$\nu_e$, from neutron decay, shown only for illustration).  We assume equal fluxes of $\nu$ and $\bar{\nu}$.  In this plot, neutrino energy is fixed at $E_\nu = 100$~TeV for illustration, but our limits are obtained using energy-averaged flavor ratios $\langle f_{\alpha,\oplus} \rangle$ (see main text), which behave similarly with $V_{e\beta}$.  For every value of $V_{e\mu}$, we scan over values of the standard mixing parameters within their $1\sigma$ ranges\ \cite{deSalas:2017kay} under normal ordering (NO).  We include the IceCube $1\sigma$ flavor contours that we use to set limits on the new interaction: the current one\ \cite{Aartsen:2015knd} (``IceCube 2015'') and projections for IceCube\ \cite{Aartsen:2017mau} (``IceCube 2017'') and IceCube-Gen2\ \cite{Aartsen:2014njl, Bustamante:2015waa}.  For comparison, we show the regions of $f_{\alpha,\oplus}$ allowed by standard mixing at $1\sigma$.}
\end{figure}

{\bf Flavor ratios at the sources.---}  We expect high-energy astrophysical neutrinos to be produced in the decay of charged pions made in $pp$ and $p\gamma$ collisions, \ie, $\pi^+ \to \mu^+ \nu_\mu \to e^+ \nu_e \bar{\nu}_\mu \nu_\mu$ and its charge-conjugate.  Thus, neutrinos leave the sources with flavor ratios $(f_{e,{\rm S}}:f_{\mu,{\rm S}}:f_{\tau,{\rm S}}) = \left( \frac{1}{3}:\frac{2}{3}:0 \right)$.  In the main text, we derive limits using this nominal expectation for $f_{\alpha,{\rm S}}$.  In the Supp.\ Mat., we consider the alternative ``muon-damped'' case $(0:1:0)_{\rm S}$, which might occur at $E_\nu \gtrsim 1$~PeV if secondary muons lose energy via synchrotron radiation before decaying, so that high-energy neutrinos come only from the direct decay of pions.  Our conclusions are unaffected by this choice.  In \figu{f_E_vary_V}, in addition to these two cases, we show, only for illustration, the case $(1:0:0)_{\rm S}$ --- a pure-$\nu_e$ flux coming, \eg, from neutron decay.


{\bf Flavor ratios at Earth.---}  At Earth, due to mixing, the ratios become $f_{\alpha,\oplus} = \sum_{\beta=e,\mu,\tau} P_{\beta\alpha} f_{\beta,{\rm S}}$.  Under standard mixing, \ie, if $V_{e\beta}$ is zero, the ratios at Earth are approximately $\left( \frac{1}{3}:\frac{1}{3}:\frac{1}{3} \right)_\oplus$.  If $V_{e\beta}$ is nonzero, the ratios at Earth depend on $g_{e\beta}^\prime$ and $m_{e\beta}^\prime$.  Since the vacuum contribution to  mixing scales $\propto 1/E_\nu$, at the energies recorded by IceCube it might be sub-dominant, making flavor ratios sensitive probes of new physics\ \cite{Barenboim:2003jm, Kashti:2005qa, Xing:2006uk, Lipari:2007su, Pakvasa:2007dc, Esmaili:2009dz, Lai:2009ke, Choubey:2009jq, Bustamante:2010nq, Mena:2014sja, Xu:2014via, Fu:2014isa, Chen:2014gxa, Palomares-Ruiz:2015mka,
Aartsen:2015ivb, Palladino:2015vna, Arguelles:2015dca, Bustamante:2015waa, Shoemaker:2015qul, Nunokawa:2016pop, Vincent:2016nut, Gonzalez-Garcia:2016gpq, Reynoso:2016hjr, Bustamante:2016ciw, Brdar:2016thq, Lai:2017bbl, Klop:2017dim, Rasmussen:2017ert, Sui:2018bbh}.

We adopt the likely scenario \cite{Murase:2013rfa, Murase:2015xka} in which the flux consists of equal parts of $\nu$ and $\bar{\nu}$, as expected from neutrino production via $pp$ collisions\ \cite{Kelner:2006tc}.  At Earth, the flavor ratios are calculated by averaging over $\nu$ and $\bar{\nu}$, since IceCube cannot distinguish between them.  

Figure \ref{fig:f_E_vary_V} shows how the flavor ratios at Earth vary with the potential.  When the potential is small, the flavor ratios are contained inside the small region expected from standard mixing\ \cite{Bustamante:2015waa}.  When the potential is large, mixing turns off and the flavor composition exits the ``theoretically palatable region'' accessible by standard mixing\ \cite{Bustamante:2015waa}.  In-between, the wiggles in the flavor ratios are due to a new resonance in the mixing parameters, driven by the long-range potential; see the Supp.\ Mat.


{\bf Flavor ratios in IceCube.---}  In IceCube, TeV--PeV astrophysical neutrinos\ \cite{Aartsen:2013bka,Aartsen:2013jdh,Aartsen:2013eka,Aartsen:2014gkd,Aartsen:2014muf,Aartsen:2015xup,Aartsen:2015knd,Aartsen:2015rwa,Aartsen:2016xlq} scatter off nucleons; scattered charged particles shower and radiate Cherenkov light that is collected by photomultipliers.  In general, it is not possible to identify flavor on an event-by-event basis\ \cite{Palomares-Ruiz:2015mka,Palladino:2015zua,Bustamante:2015waa}, but it is possible to infer the flavor ratios of the astrophysical flux by comparing relative numbers of different event classes\ \cite{Mena:2014sja, Aartsen:2015ivb, Aartsen:2015knd, Palomares-Ruiz:2015mka, Vincent:2016nut, Li:2016kra}.

Figure \ref{fig:f_E_vary_V} shows the latest published IceCube flavor results at $1\sigma$~C.L.\ \cite{Aartsen:2015knd};  the best-fit composition is $(0.49:0.51:0)_\oplus$.  Presently, the nominal expectation $\left( \frac{1}{3}:\frac{1}{3}:\frac{1}{3} \right)_\oplus$ is $\sim$$1\sigma$ removed from the best fit\ \cite{Aartsen:2015knd}.  Below, we explore also projections where the IceCube best-fit point moves closer to the nominal expectation.  At confidence levels higher than $1\sigma$, present IceCube contours are significantly wider\ \cite{Aartsen:2015knd}.  Present IceCube results disfavor a scenario without oscillations --- where $f_{\alpha,\oplus} = f_{\alpha,{\rm S}}$ --- at $\sim$$1\sigma$, which allows us to constrain the new interaction at this level.  Figure \ref{fig:f_E_vary_V} also shows a preliminary update of the IceCube flavor sensitivity\ \cite{Aartsen:2017mau}, and an estimate\ \cite{Bustamante:2015waa} for the IceCube-Gen2 upgrade\ \cite{Aartsen:2014njl}.  Both are artificially centered on the nominal expectation for $f_{\alpha,\oplus}$.

Before contrasting our flavor predictions with IceCube results, we fold in the neutrino energy spectrum.  The incoming flux of $\nu_\alpha+\bar{\nu}_\alpha$ is $\Phi_\alpha(E_\nu) \propto f_{\alpha, \oplus}(E_\nu) \cdot E_\nu^{-\gamma}$.  Different analyses yielded different values of the spectral index: $\gamma = 2.50$, using events of all classes\ \cite{Aartsen:2015knd}, and $\gamma = 2.13$, using only upward-going muons\ \cite{Aartsen:2016xlq}.  Below, we consider these two possibilities; the choice has little effect.
The average flux in the interval 25~\text{TeV}--2.8~\text{PeV}\ \cite{Aartsen:2015knd}, where the IceCube flavor results apply, is $\langle \Phi_\alpha \rangle \approx (2.8~\text{PeV})^{-1} \int dE_\nu ~\Phi_\alpha(E_\nu)$.  From this, we define energy-averaged ratios $\langle f_{\alpha,\oplus} \rangle \equiv \langle \Phi_\alpha \rangle / \sum_\beta \langle \Phi_\beta \rangle$, our observables.  The behavior of $\langle f_{\alpha,\oplus} \rangle$ resembles that of $f_{\alpha,\oplus}$ in \figu{f_E_vary_V}.


{\bf Limit-setting procedure.---}  To constrain the $Z_{e\beta}^\prime$, we compare $\langle f_{\alpha,\oplus} \rangle$ to the IceCube flavor measurements.  This way, the IceCube analysis systematics involved in extracting the flavor ratios are already implicitly taken into account.  We describe our procedure below.

For a particular choice of values $(m_{e\beta}^\prime, g_{e\beta}^\prime)$, we independently vary the standard mixing parameters $\theta_{12}$, $\theta_{23}$, $\theta_{13}$, $\delta_{\rm CP}$, $\Delta m^2_{21}$, and $\Delta m^2_{31}$ within their experimentally allowed $1\sigma$ ranges, on a fine grid.  We use the ranges from \Ref\ \cite{deSalas:2017kay}, assuming a normal neutrino mass ordering, which is currently favored over the inverted one at $3.5\sigma$\ \cite{deSalas:2018bym}.  Later, we comment on the inverted ordering.  For each choice of values of the mixing parameters, we compute the energy-averaged ratios $(\langle f_{e,\oplus} \rangle : \langle f_{\mu,\oplus} \rangle : \langle f_{\tau,\oplus} \rangle)$.  We impose a simple hard cut: if the ratios calculated for all choices of values of the mixing parameters fall outside the $1\sigma$ IceCube contour, then the point $(m_{e\beta}^\prime, g_{e\beta}^\prime)$ is disfavored at, at least, $1\sigma$~C.L.  Otherwise, the point $(m_{e\beta}^\prime, g_{e\beta}^\prime)$ is allowed.  We scan $m_{e\beta}^\prime$ and $g_{e\beta}^\prime$ over wide intervals and repeat the above procedure for every value.

We also derive limits based on the projected IceCube and IceCube-Gen2 flavor contours in \figu{f_E_vary_V}.  Even though by the time of completion of IceCube-Gen2 --- late 2020s --- mixing parameters should be known to higher precision\ \cite{Coloma:2012wq}, we have tested that already now their uncertainty is not a limiting factor.  Using reduced uncertainties --- $5\%$ for $\delta_{\rm CP}$ and $1\%$ for all other parameters --- projected limits are only slightly better.


{\bf Results.---}  Figure \ref{fig:limits_emu} shows that our limits on the coupling $g_{e\mu}^\prime$ are the strongest for masses below $10^{-18}$~eV.  The limits on $g_{e\tau}^\prime$ are similar.  They are in the Supp.\ Mat., which contains also limits for alternative choices.

Using current IceCube flavor results, we can place an upper limit because the no-oscillation point $\left( \frac{1}{3}:\frac{2}{3}:0 \right)_\oplus$ --- reachable with large couplings --- lies outside the IceCube contour; see \figu{f_E_vary_V}.  We can place a lower limit too because the standard-mixing region --- reachable with small couplings --- also lies outside the contour. 

Figure \ref{fig:limits_emu} also shows limits derived using the projected IceCube and IceCube-Gen2 flavor contours.  Both contours fully contain the standard-mixing region, but not $\left( \frac{1}{3}:\frac{2}{3}:0 \right)_\oplus$; see \figu{f_E_vary_V}.  Hence, in these projections, we can set only upper limits.  With IceCube-Gen2, limits could be 4 times better than the current ones.

Our limits are robust against uncertainties in the shape of the neutrino spectrum and choice of mass ordering.  Soft ($\gamma = 2.50$) and hard ($\gamma = 2.13$) spectra yield marginally different limits, since the energy-averaged $\langle f_{\alpha,\oplus} \rangle$ are dominated by low energies; we show results only for $\gamma = 2.50$.  For the alternative choice $\left(0:1:0\right)_{\rm S}$, the limits improve by a factor of 2.5--5, depending on $m_{e\mu}^\prime$.  Switching to inverted mass ordering has little effect on the upper limits, since the no-oscillation point still lies outside the $1\sigma$ flavor contour.  However, the lower limits derived using current IceCube flavor results deteriorate, on account of our hard $1\sigma$ cut, because most of the standard-mixing region now falls inside the IceCube contour, thus allowing smaller values of the coupling. 

Our limits outperform existing ones.  Existing direct limits come from atmospheric\ \cite{Joshipura:2003jh}, and solar and reactor neutrinos\ \cite{GonzalezGarcia:2006vp, Bandyopadhyay:2006uh}.  Indirect limits come from tests of non-standard neutrino interactions\ \cite{Mitsuka:2011ty, Ohlsson:2012kf, Gonzalez-Garcia:2013usa} --- calculated for \figu{limits_emu} following \Ref\ \cite{Wise:2018rnb}, but only up to $m_{e\beta}^{-1} = R_\oplus$ and using our long-range potential --- tests of the equivalence principle\ \cite{Schlamminger:2007ht} and fifth force\ \cite{Adelberger:2009zz}, black-hole superradiance\ \cite{Baryakhtar:2017ngi}, and stellar cooling\ \cite{Hardy:2016kme}.  Figure \ref{fig:limits_emu} shows the most competitive limits; for a full review, including collider limits at higher masses, see \Ref\ \cite{Wise:2018rnb}.


{\bf Limitations and improvements.---}  The main factor limiting our sensitivity is the uncertainty in flavor measurements.  However, it is expected to improve in the near future: a larger neutrino event sample and advances in flavor reconstruction\ \cite{Aartsen:2018vez} will tighten the IceCube flavor results.  This will allow the extracted limits to have a higher statistical significance.  New directions in flavor-tagging techniques --- \eg, muon and neutron echoes\ \cite{Li:2016kra} --- could aid.  Proposals to distinguish $\bar{\nu}$ from $\nu$ could test our assumption of equal fluxes of each\ \cite{Anchordoqui:2004eb, Bhattacharya:2011qu, Barger:2014iua}.  

If the relic neutrino background contains equal numbers of $\nu_e$ and $\bar{\nu}_e$, it may partially screen out the long-range potential sourced by distant electrons\ \cite{Dolgov:1995hc, Blinnikov:1995kp, Joshipura:2003jh, Grifols:2003gy}.
We have not considered this effect in our calculation, but it would exclusively affect the sensitivity to couplings $g_{e\beta}^\prime \lesssim 10^{-29}$, \ie, the sensitivity due to cosmological electrons.   For those couplings, the distance at which this effect becomes relevant --- the Debye length\ \cite{Joshipura:2003jh} --- is roughly a factor-of-10 smaller than the interaction range $1/m_{e\beta}^\prime$ to which we are sensitive, given by the values along the curve in \figu{potential}.


{\bf Summary.---}  In extending the Standard Model (SM), large-scale neutrino telescopes --- IceCube and future IceCube-Gen2 and KM3NeT\ \cite{Adrian-Martinez:2016fdl} --- provide valuable guidance\ \cite{Ahlers:2018mkf}, thanks to their detection of neutrinos with the highest energies.
We searched for new long-range neutrino-electron interactions, mediated by ultra-light mediators, via the flavor composition of high-energy astrophysical neutrinos in IceCube.  For the first time, we reached the ultimate sensitivity to these interactions, as a result of using the highest neutrino energies  and accounting for the huge number of electrons in the local and distant Universe.  Our results, the strongest to date, disfavor the existence of long-range neutrino-electron interactions, crucially complementing results from collider searches for new short-range interactions.


\smallskip
{\bf Acknowledgements.} MB is supported by the Danmarks Grundforskningsfond Grant 1041811001.  SKA is supported by DST/INSPIRE Research Grant IFA-PH-12, Department of Science and Technology, India and the Young Scientist Project INSA/SP/YSP/144/2017/1578 from the Indian National Science Academy.  We thank Atri Bhattacharya, Peter Denton, Andr\'e de Gouv\^{e}a, Yasaman Farzan, Matheus Hostert, Shirley Li, Subhendra Mohanty, and Subir Sarkar for feedback and discussion.  This work used resources provided by the High Performance Computing Center at the University of Copenhagen.  We acknowledge the use of the {\tt python-ternary} package by Marc Harper {\it et al.} to produce ternary plots.




%


\newpage
\clearpage

\appendix


\onecolumngrid

\begin{center}
 \large
 Supplemental Material for\\
 \smallskip
 {\it A Universe's worth of electrons to probe long-range interactions\\ of high-energy astrophysical neutrinos}
\end{center}

\twocolumngrid


\section{Derivation of the long-range potential}
\label{appendix:potential_derivation}

Due to the $L_e-L_\beta$ ($\beta = \mu, \tau$) symmetry, an electron sources a Yukawa potential
\begin{equation}\label{equ:potential_one_electron}
 V_{e\beta} = - \frac{ g_{e\beta}^{\prime 2} } { 4\pi d } e^{-m_{e\beta}^\prime d}
\end{equation}
at a distance $d$ from it, where $g_{e\beta}^\prime$ is the new coupling between electrons and neutrinos, and $m_{e\beta}^\prime$ is the mass of the $Z_{e\beta}^\prime$ that acts as mediator.  For a given value of the mass, the range of the interaction is $1/m_{e\beta}^\prime$; beyond that, the potential is exponentially suppressed.

Because we focus on tiny mediator masses, the interaction range is between meters and thousands of Gpc.  Below, we compute the most important contributions to the potential, coming from electrons in the Earth, Moon, Sun, Milky Way, and cosmological electrons.  When calculating the number of electrons $N_e$ in a concentration of matter, we assume that the matter is isoscalar --- it has roughly equal number of protons $N_p$ and neutrons $N_n$ --- and electrically neutral, so that the electron fraction in them is $Y_e \equiv N_e / (N_p + N_n) = 0.5$.  With this, we convert from baryon density to electron density.


\subsection{Electrons in the Earth}

To calculate the potential due to the $N_{e,\oplus} \sim 4 \cdot 10^{51}$ electrons inside the Earth, we compute the electron column densities traversed by neutrinos inside the Earth prior to arriving at IceCube.  To do this, we use the profile of electron number density $n_{e,\oplus}$ built from the matter density profile of the Preliminary Reference Earth Model (PREM)\ \cite{Dziewonski:1981xy}.  The profile, constructed from seismic data, consists in concentric layers of increasing density towards the center of the Earth.

At the position of IceCube, the net potential acting on neutrinos arriving from all directions is
\begin{eqnarray}
 V_{e\beta}^\oplus &=&
 2 \pi \frac{g_{e\beta}^{\prime 2}}{4\pi} \int_0^\pi d\theta \int_0^{r_{\max}(\theta)} dr
 ~r ~\langle n_{e,\oplus}(r,\theta) \rangle_\theta \nonumber \\
 && \qquad\qquad\qquad\qquad\qquad \times \sin \theta ~e^{-m_{e\beta}^\prime r} \;,
\end{eqnarray}
where $R_\oplus = 6371$~km is the radius of the Earth, $\langle n_{e,\oplus} \rangle_\theta$ is the average electron density along the direction given by $\theta$, and
$r_{\max}(\theta) = (R_\oplus-d_{\rm IC}) \cos \theta + \left[ (R_\oplus-d_{\rm IC})^2 \cos^2 \theta + (2R_\oplus-d_{\rm IC}) d_{\rm IC} \right]^{1/2}$ is the length of the chord traversed by the neutrino inside the Earth, with $d_{\rm IC} = 1.5$~km the approximate depth of IceCube.

To compute the potential due to standard matter effects inside the Earth, we adopt a simpler prescription: $V_{\rm mat}^\oplus = \sqrt{2} G_F \langle n_e^\oplus \rangle$, where $\langle n_e^\oplus \rangle \equiv Y_e \langle n_N \rangle / (2 m_p)$ is the average electron density and $\langle n_N \rangle \approx 5.5$~g~cm$^{-3}$ is the average nucleon density according to the PREM.  We do this because, in the regime where standard matter effects become important --- when the interaction range is smaller than $R_\oplus$ --- other limits on $g_{e\beta}^\prime$ are stronger, as shown in \figu{limits_emu}, avoiding the need for a more sophisticated calculation.


\subsection{Electrons in the Moon and the Sun}

We treat the Moon and the Sun as point sources of electrons.  The potential $V_{e\beta}^{\leftmoon}$ due to electrons in the Moon is obtained by evaluating \equ{potential_one_electron} at $d = d_{\leftmoon} \approx 4 \cdot 10^5$~km --- the distance between the Earth and the Moon --- and multiplying it by $N_{e,\leftmoon} \sim 5 \cdot 10^{49}$ --- the number of electrons in the Moon.  Similarly, the potential $V_{e\beta}^{\astrosun}$ due to electrons in the Sun is obtained by evaluating \equ{potential_one_electron} at $d = d_{\astrosun} =1$~A.U. --- the distance between the Earth and the Sun --- and multiplying it by $N_{e,\astrosun} \sim 10^{57}$ --- the number of electrons in the Sun.


\subsection{Electrons in the Milky Way}

\setcounter{figure}{0}
\renewcommand{\thefigure}{A\arabic{figure}}
\begin{figure}[t!]
 \centering
 \includegraphics[width=\columnwidth]{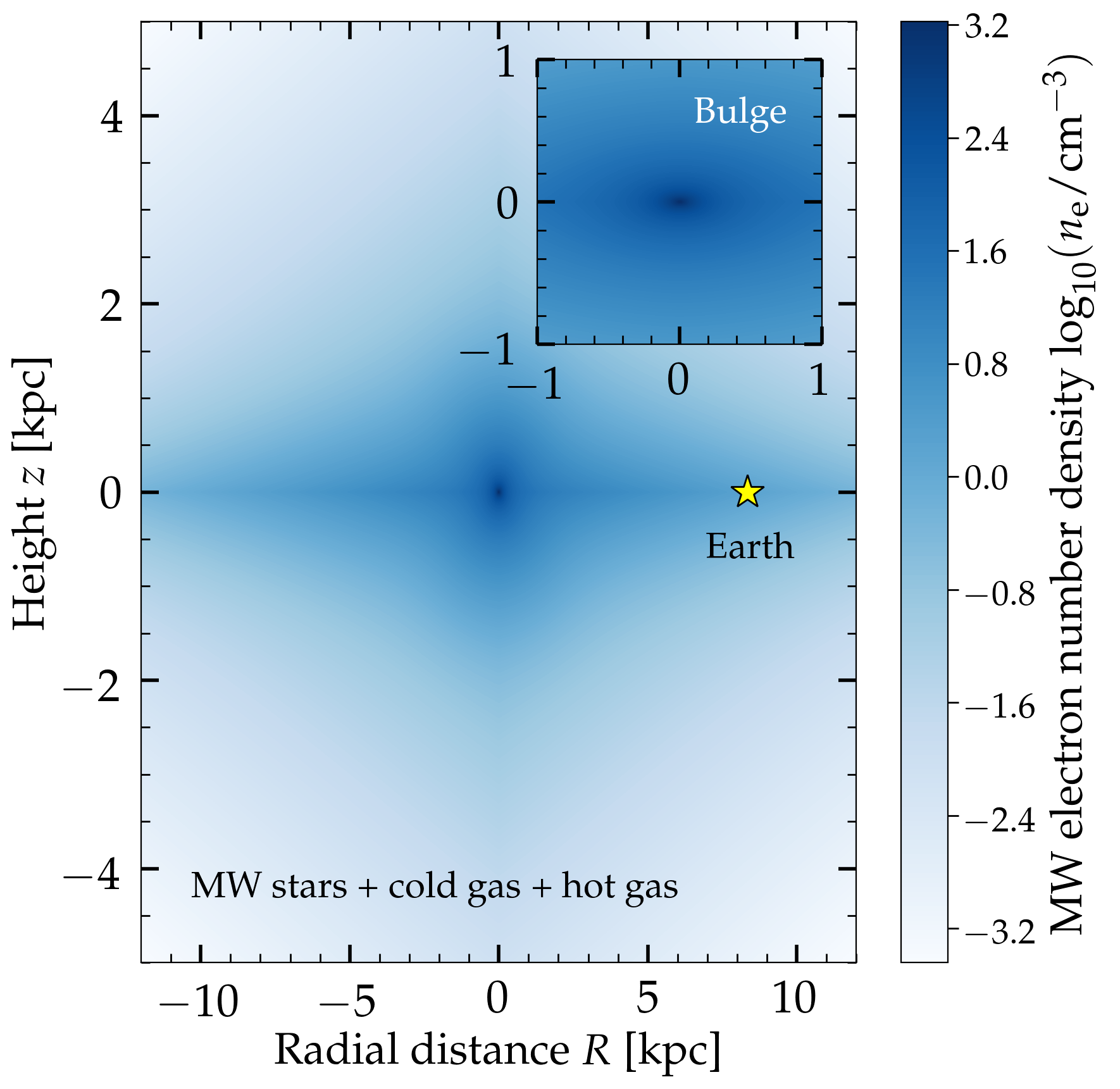}
 \caption{\label{fig:mw_electron_number_density}Density of electrons in the Milky Way, in Galactocentric coordinates.  Electrons are distributed in the central bulge, thin disc, and thick disc of stars and cold gas\ \cite{McMillan:2011wd}, and in the diffuse halo of hot gas\ \cite{Miller:2013nza}.}
\end{figure}

The baryonic content of the Milky Way consists of stars and cold gas --- distributed in a central bulge, a thick disc, and a thin disc --- and hot gas --- distributed in a diffuse halo.  We compute the potential due to the total $N_{e,{\rm MW}} \sim 10^{67}$ electrons, assuming, as before, $Y_e = 0.5$.  

Figure \ref{fig:mw_electron_number_density} shows the density of electrons in the Milky Way.  For the central bulge, thick disc, and thin disc, we assume the simplified profiles of matter density from \Ref\ \cite{McMillan:2011wd}.  These were obtained via a Bayesian fit to photometric and kinematic data.  Each of the three components is modeled as a flat cylinder centered on the Galactic Center, with the matter density exponentially falling away from the axis and from the Galactic Plane.  We adopt the parameter values from the ``convenient model'' of \Ref\ \cite{McMillan:2011wd}.  For the diffuse halo of hot gas, we assume the spherical saturated matter density profile from \Ref\ \cite{Miller:2013nza}, obtained from measurements of O VII K$\alpha$ x-ray absorption lines using XMM-Newton.  The density is highest at the Galactic Center and falls exponentially outwards.  

We calculate the potential due to Milky Way electrons by integrating the electron column density along all incoming neutrino directions, \ie,
\begin{eqnarray}\label{equ:potential_mw}
 V_{e\beta}^{\rm MW} &=&
 \frac{g_{e\beta}^{\prime 2}}{4\pi} \int_0^\infty dr \int_0^\pi d\theta \int_0^{2\pi} d\phi 
 ~r ~n_{e, {\rm MW}}(r,\theta,\phi) \nonumber \\
 && \qquad\qquad\qquad\qquad\qquad \times \sin\theta ~e^{-m_{e\beta}^\prime r} \;,
\end{eqnarray}
with the coordinate system centered at the position of the Earth, which is located 8.33~kpc away from the Galactic Center\ \cite{McMillan:2011wd}. 
The potential is dominated by electrons in stars and cold gas.  Though the halo of hot gas accounts for a significant fraction of the baryonic content of the Milky Way, its density is low, so halo electrons are only a tiny contribution to the total potential in \equ{potential_mw}.


\subsection{Cosmological electrons}

\renewcommand{\thefigure}{A\arabic{figure}}
\begin{figure}[t!]
 \centering
 \includegraphics[width=0.49\textwidth]{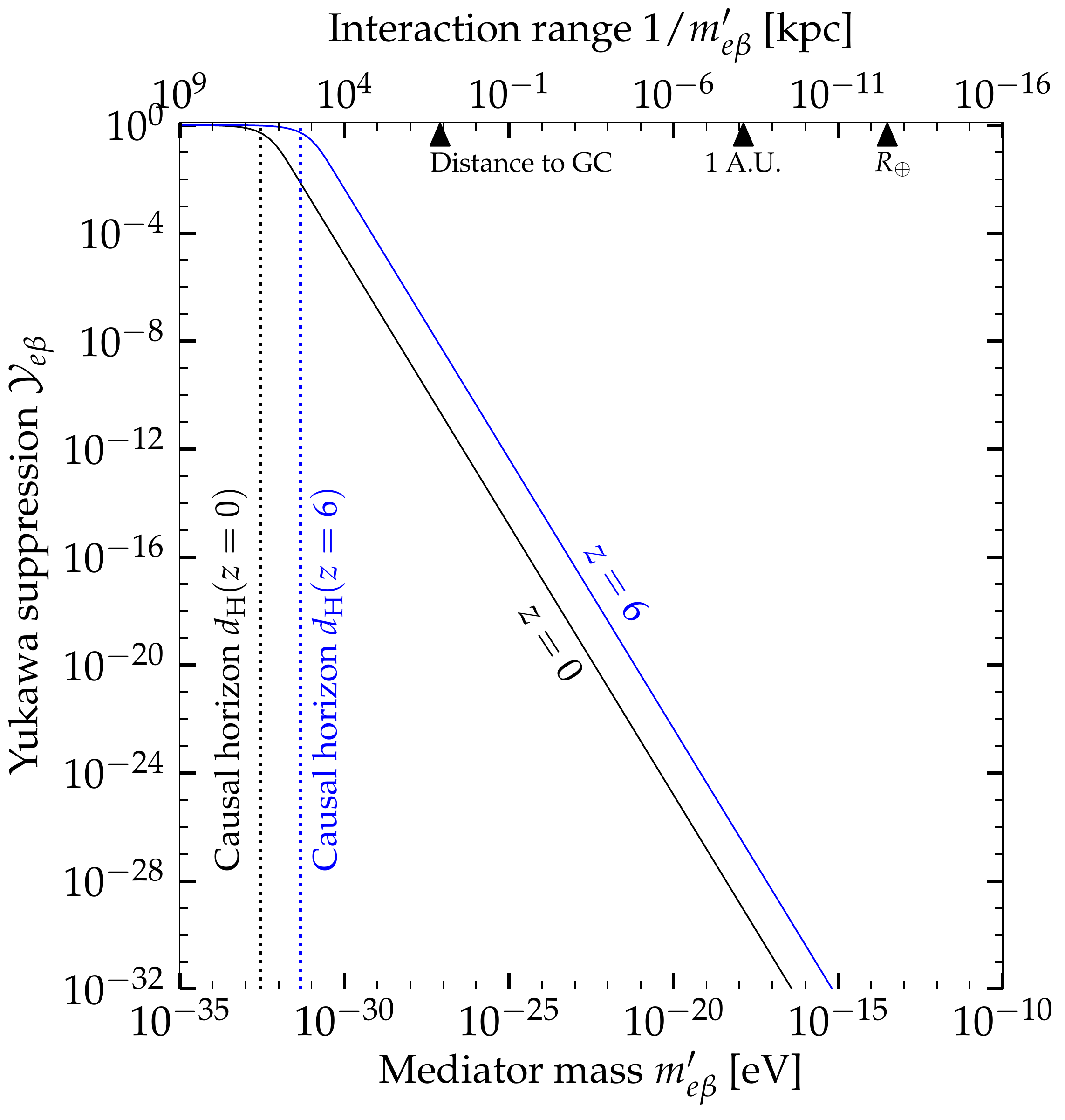}
 \caption{\label{fig:yukawa_suppression}Yukawa suppression $\mathcal{Y}_{e\beta}$ of the potential due to cosmological electrons, as a function of mediator mass $m_{e\beta}^\prime$, for two fixed values of redshift: $z=0$ and $z=6$.  For comparison, we show the causal horizon for the two choices.}
\end{figure}

In addition to the electron repositories in the local Universe, there is, at all redshifts, a cosmological distribution of electrons.  The huge number of cosmological electrons --- $N_{e,{\rm cos}} \sim 10^{79}$ --- is what allows us to set the best bounds on the coupling $g_{e\beta}^\prime$ at the lowest values of mediator mass, where the interaction range is of the order of the size of Universe, or larger.  Below, we calculate the potential due to cosmological electrons.

Consider a neutrino that sits at the center of a sphere of radius $R$ that is homogeneously filled with a constant number density $n_e$ of electrons.  The integrated long-range potential at the position of the neutrino is then
\begin{equation}\label{equ:potential_def_general}
 V_{e\beta} = 
 g_{e\beta}^{\prime 2} n_e
 \left[ 
 \frac { 1-e^{-m_{e\beta}^\prime R}(1+m_{e\beta}^\prime R) } { m_{e\beta}^{\prime 2} } 
 \right] \;.
\end{equation}

IceCube neutrinos are predominantly extragalactic, and presumably generated in sources at different redshifts.  Because of the cosmological expansion, the density of cosmological electrons and the potential that they source varies with redshift.  We take into account these effects as follows.

The causal horizon defines the largest possible region within which events can be causally connected to each other\ \cite{Weinberg:2008zzc}.  At redshift $z$, the comoving size of the causal horizon centered around the neutrino is
\begin{equation}
 d_{\rm H} \left( z \right) = H_0^{-1} \int_0^{\left(1+z\right)^{-1}} \frac{dx}{h\left(x\right)} \;,
\end{equation}
where $H_0 = 100 h$ km s$^{-1}$ Mpc$^{-1}$ is the Hubble constant, with $h = 0.673$~\cite{Agashe:2014kda}, $x \equiv \left( 1+z \right)^{-1}$, and $h\left(x\right) \equiv H\left(x\right) / H_0$, with the Hubble parameter
$H\left(x\right) = H_0 x \sqrt{ \Omega_\Lambda^0 x^2 + \Omega_\text{M}^0 x^{-1} }$.  We adopt a $\Lambda$CDM cosmology with vacuum energy density $\Omega_\Lambda = 0.692$ and matter density $\Omega_{\rm M} = 0.308$\ \cite{Ade:2015xua}.  The causal horizon changes from about 14.5 Gpc at $z=0$ to about 0.9 Gpc at $z=6$.

\setcounter{figure}{0}
\renewcommand{\thefigure}{B\arabic{figure}}
\begin{figure*}[t!]
 \centering
 \includegraphics[width=0.49\textwidth]{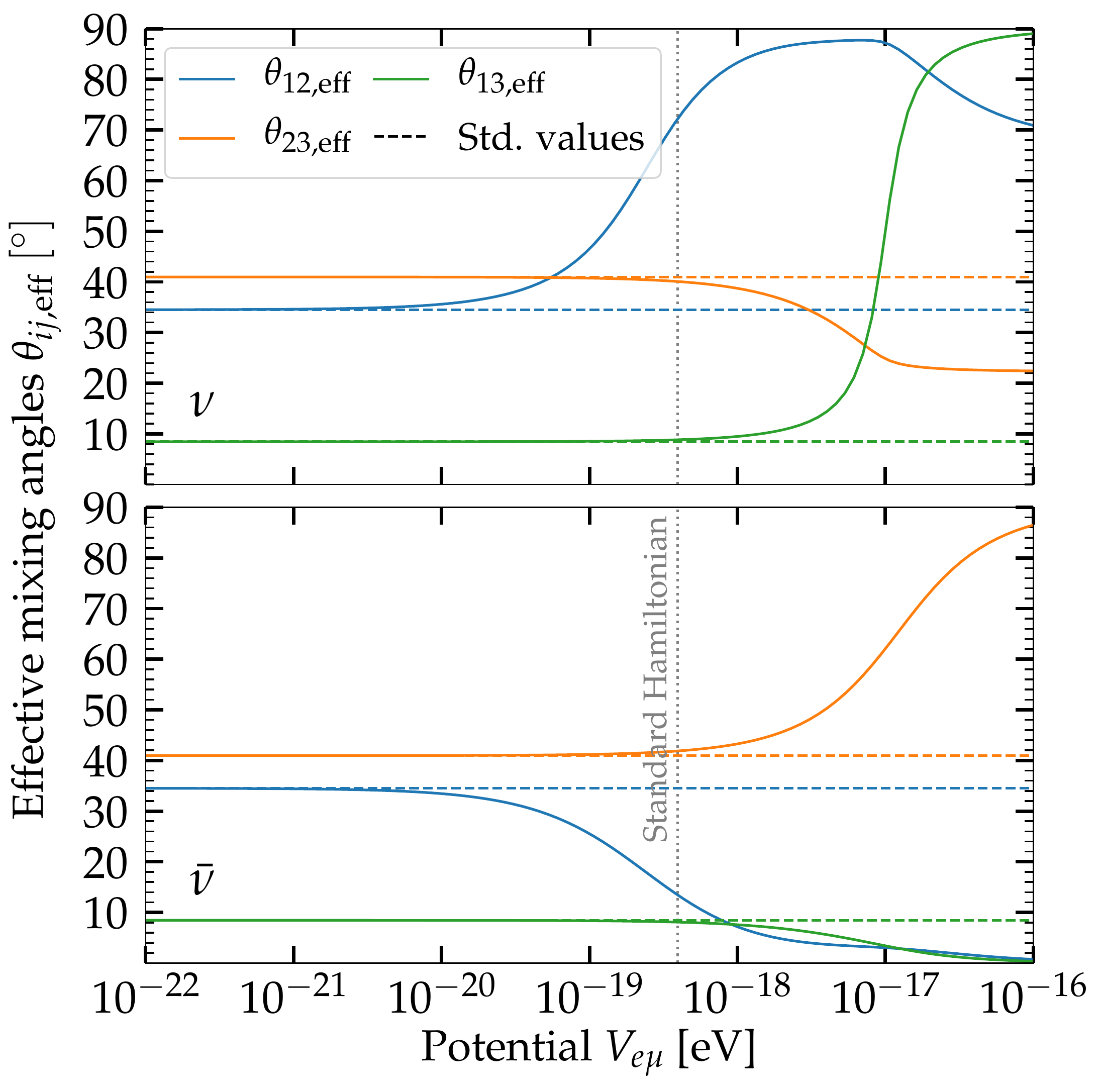}
 \includegraphics[width=0.49\textwidth]{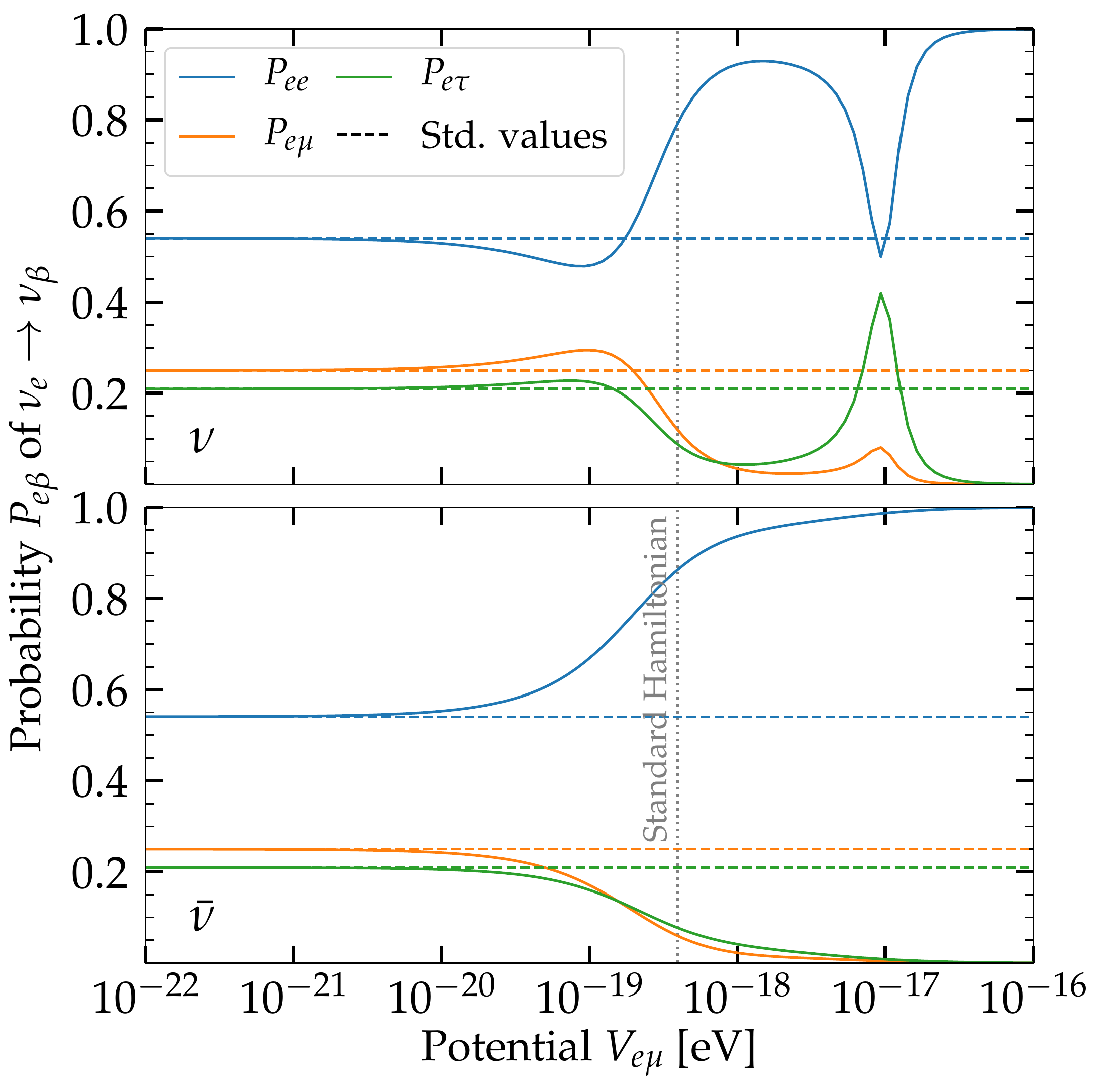}
 \caption{\label{fig:mixing_prob_lrp}Effective neutrino mixing parameters ({\it left}) and modified probabilities $P_{ee}$, $P_{e\mu}$, and $P_{e\tau}$ ({\it right}), in the presence of the new long-range interaction from the $L_e-L_\mu$ symmetry, at neutrino energy $E_\nu = 100$~TeV, as a function of the potential $V_{e\mu}$.  For comparison, we show the value of the $ee$ element of $\matr{H}_{\rm vac}$ at this energy.  Standard mixing parameters are fixed to their best-fit values under normal mass ordering from \Ref\ \cite{deSalas:2017kay}.  Dashed lines show the standard values of the quantities, \ie, for $V_{e\mu} = 0$.  Top panels are for neutrinos; bottom panels are for anti-neutrinos.  When computing limits, we consider equal fluxes of $\nu$ and $\bar{\nu}$.}
\end{figure*}

The content of baryonic matter inside the causal horizon (see \eq~(16.105) in \Ref~\cite{Giunti:2007ry}) is
\begin{equation}
 M_{\rm H} \left( z \right)
 = \frac{H_0^2}{16 G_N} d_{\rm H}^3 \left( z \right) \Omega_b^0 \;,
\end{equation}
where $\Omega_b^0 \approx 0.02207 h^{-2} \approx 0.05$~\cite{Agashe:2014kda} is the density of baryons in the local Universe.  The total mass is predominantly made up of protons, neutrons, and electrons, \ie, $M_{\rm H} \left( z \right) \simeq N_p \left( z \right) m_p + N_n \left( z \right) m_n + N_e \left( z \right) m_e$, where $m_p$, $m_n$, and $m_e$ are the masses of one proton, neutron, and electron.  We estimate the number of electrons by assuming that the number of protons and neutrons is roughly equal ($N_p \approx N_n$) and the net electric charge is zero ($N_p \approx N_e$). Taking $m_n \approx m_p$, this results in 
\begin{equation}
 N_e \left( z \right) \simeq M_{\rm H} \left( z \right) / \left( 2 m_p + m_e \right)  \;.
\end{equation}
By evaluating \equ{potential_def_general} with $R = d_{\rm H}(z)$ and $n_e = N_e(z) / V_{\rm H}(z)$, with $V_{\rm H}(z) \equiv (4/3) \pi d_{\rm H}^3(z)$ the causal volume, the potential acting on a neutrino at redshift $z$ is
\begin{equation}\label{equ:potential_def}
 V_{e\beta}^{\rm cos}(z) = \mathcal{C}_{e\beta}(z) \cdot \mathcal{Y}_{e\beta}(z) \;.
\end{equation}
The term due to the Coulomb part of the potential,
\begin{equation}\label{equ:potential_def_coulomb}
 \mathcal{C}_{e\beta}(z) 
 = \frac{3}{2} \frac{g_{e\beta}^{\prime 2}}{4\pi} \frac{N_e(z)}{d_{\rm H}(z)} \;,
\end{equation}
describes a potential with infinite range, mediated by a massless mediator.  The Yukawa suppression,
\begin{equation}\label{equ:potential_def_yukawa}
 \mathcal{Y}_{e\beta}(z) 
 = \frac{2}{[m_{e\beta}^\prime d_{\rm H}(z)]^2} \left\{ 1 - e^{-m_{e\beta}^\prime d_{\rm H}(z)} [1+m_{e\beta}^\prime d_{\rm H}(z)] \right\} \;.
\end{equation}
reflects the reduced interaction range due to the mediator being massive and the finite size of the causal horizon.  Smaller values of $\mathcal{Y}_{e\beta}$ represent stronger suppression.

Figure \ref{fig:yukawa_suppression} illustrates the behavior of the Yukawa suppression.  For a fixed redshift, the suppression is important --- \ie, $\mathcal{Y}_{e\beta} \ll 1$ --- as long as the interaction range $1/m_{e\beta}^\prime$ is small compared to the causal horizon.  This means that the contribution of electrons located far from the neutrino is exponentially suppressed.  This occurs for $m_{e\beta}^\prime \gtrsim 10^{-31}$~eV at $z=6$ and $m_{e\beta}^\prime \gtrsim 10^{-33}$~eV at $z=0$.  On the other hand, if the range is comparable to or larger than the causal horizon, there is no Yukawa suppression, \ie, $\mathcal{Y}_{e\beta} \approx 1$.  In this case, the interaction range is effectively infinite, that is, larger than the size of the causally connected Universe.


\section{Flavor mixing in a long-range potential}
\label{appendix:flavor_mixing_lrp}

In the presence of the long-range potential, the average flavor-transition probability is $P_{\alpha\beta}(E_\nu) = \sum_{i=1}^3 \lvert U^\prime_{\alpha i}(E_\nu) \rvert^2 \lvert U^\prime_{\beta i}(E_\nu) \rvert^2$, where $\matr{U}^\prime$ is the matrix that diagonalizes the total Hamiltonian $\matr{H}_{e\beta}(E_\nu,g_{e\beta}^\prime, m_{e\beta}^\prime) \equiv \matr{H}_{\rm vac}(E_\nu) + \matr{V}_{e\beta}(g_{e\beta}^\prime, m_{e\beta}^\prime) + \Theta(R_\oplus-m_{e\beta}^{\prime -1}) \matr{V}_{\rm mat}^\oplus$.  The new interaction between neutrinos and electrons modifies the effective mixing angles $\theta_{12,{\rm eff}}$, $\theta_{13,{\rm eff}}$, and $\theta_{23,{\rm eff}}$, and the effective squared-mass differences $\Delta m_{21,{\rm eff}}^2$ and $\Delta m_{31,{\rm eff}}^2$.  The effective mixing angles are identified by writing the $\matr{U}^\prime$ as a PMNS-like matrix, while the effective squared-mass differences are the eigenvalues of $\matr{H}_{e\beta}$.

Standard flavor mixing occurs because the neutrino flavor and mass bases are different, \ie, because $\matr{H}_{\rm vac}$ is non-diagonal.  Indeed, if $\matr{V}_{e\beta} \ll \matr{H}_{\rm vac} + \Theta(R_\oplus-m_{e\beta}^{\prime -1}) \matr{V}_{\rm mat}^\oplus$, we recover standard mixing.  In \figu{potential}, this happens below the iso-contours of $V_{e\beta} = [\matr{H}_{\rm vac}(E_\nu)]_{ee} + V_{\rm mat}^\oplus$.  If, on the other hand, $\matr{V}_{e\beta} \gg \matr{H}_{\rm vac} + \Theta(R_\oplus-m_{e\beta}^{\prime -1}) \matr{V}_{\rm mat}^\oplus$, the total Hamiltonian becomes effectively diagonal and mixing turns off, \ie, $P_{\alpha\alpha} \approx 1$.  In \figu{potential}, this happens above the iso-contours.  In-between, when $\matr{V}_{e\beta} \approx \matr{H}_{\rm vac} + \Theta(R_\oplus-m_{e\beta}^{\prime -1}) \matr{V}_{\rm mat}^\oplus$, flavor mixing occurs with modified probabilities.

Figure \ref{fig:mixing_prob_lrp} shows how the effective mixing angles and probability $P_{e\beta}$ ($\beta = e, \mu, \tau$), calculated assuming the $L_e-L_\mu$ symmetry, vary with $V_{e\mu}$.  The long-range interaction induces a new resonance in the mixing of neutrinos, at $V_{e\mu} \sim 10^{-17}$~eV, on account of the potential term and the vacuum term having opposite signs.  For anti-neutrinos, this does not occur and hence the resonance is not present.  The resonance accounts for the wiggles seen in the flavor ratios in \figu{f_E_vary_V}.  Because, in obtaining our limits, we averaged over equal fluxes of $\nu$ and $\bar{\nu}$, the wiggles are damped in \figu{f_E_vary_V}.  The resonance is softer and broader in the $P_{\mu\beta}$ channels (not shown).  At higher values of the potential, mixing turns off, \ie, $P_{ee} \approx 1$.

Figure \ref{fig:mixing_prob_lrp} uses the best-fit values of the mixing angles under the normal mass ordering.  Under the inverted mass ordering (not shown), results are similar, but the curves for $P_{e\mu}$ and $P_{e\tau}$ are swapped below the resonance, though $P_{e\tau}$ remains larger than $P_{e\mu}$ at the resonance.  For the $L_e-L_\tau$ symmetry (not shown), results are similar, but $P_{e\mu}$ and $P_{e\tau}$ are swapped near the resonance.


\setcounter{figure}{0}
\renewcommand{\thefigure}{C\arabic{figure}}
\begin{figure}[t!]
 \centering
 \includegraphics[width=\columnwidth]{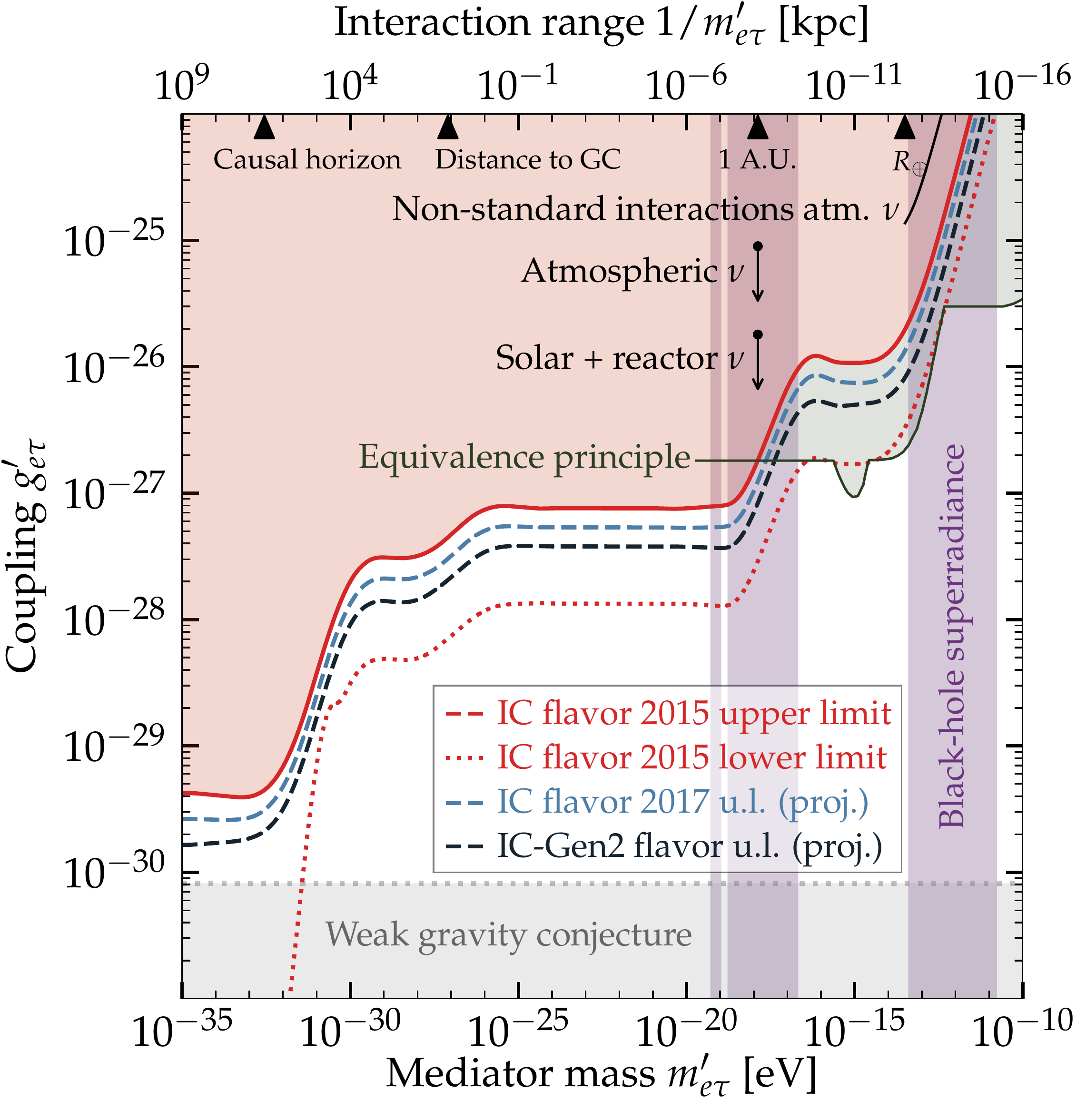}
 \caption{\label{fig:limits_etau}Same as \figu{limits_emu}, but for the $L_e-L_\tau$ symmetry.  Like in \figu{limits_emu}, we assume flavor ratios at the source $\left(\frac{1}{3}:\frac{2}{3}:0\right)_{\rm S}$ and normal mass hierarchy.}
\end{figure}

\vspace*{0.19cm}

\section{Constraints for $L_e-L_\tau$}

Figure \ref{fig:limits_etau} shows present and future constraints on $g_{e\tau}^\prime$, in analogy to \figu{limits_emu} for $g_{e\mu}^\prime$ in the main text.  The only difference compared to \figu{limits_emu} is that slightly larger values of $g_{e\tau}^\prime$ are allowed than for $g_{e\mu}^\prime$.

The similarity between the limits on $g_{e\mu}^\prime$ and $g_{e\tau}^\prime$ is evident from inspecting the behavior of $f_{\alpha,\oplus}$ as a function of the long-range potential, shown in the top row of \figu{ternary_ordering_compare}.  For both $L_e-L_\mu$ and $L_e-L_\tau$, the standard-mixing region and the point $\left(\frac{1}{3}:\frac{2}{3}:0\right)_{\oplus}$ lie outside the $1\sigma$ IceCube contour.  The main difference between the two cases is the direction of the wiggle in $f_{\alpha,\oplus}$ due to the new resonance; see Appendix~\ref{appendix:flavor_mixing_lrp}.  The similarity in the limits holds also when the inverted mass hierarchy is assumed; see Appendix \ref{appendix:mass_ordering}.


\section{The effect of mass ordering}
\label{appendix:mass_ordering}

To derive the limits on $g_{e\mu}^\prime$ in \figu{limits_emu} and on $g_{e\tau}^\prime$ in \figu{limits_etau}, we varied the standard neutrino mixing parameters $\theta_{12}$, $\theta_{23}$, $\theta_{13}$, $\delta_{\rm CP}$, $\Delta m_{21}^2$, and $\Delta m_{31}^2$ within their allowed $1\sigma$~C.L. ranges obtained from the global oscillation analysis of \Ref\ \cite{deSalas:2017kay}, assuming a normal mass ordering.  Here we explore how the limits on $g_{e\mu}^\prime$ and $g_{e\tau}^\prime$ change when we assume instead an inverted ordering.

Figure \ref{fig:ternary_ordering_compare} shows the flavor ratios at Earth $f_{\alpha,\oplus}$, evaluated at $E_\nu = 100$~TeV, for the lepton-number symmetries $L_e-L_\mu$ and $L_e-L_\tau$, and the normal and inverted mass orderings.  The top left panel is the same as \figu{f_E_vary_V}, and is reproduced here to facilitate the comparison.

Figure \ref{fig:limits_ordering_compare} shows that, using the present IceCube flavor results, switching to inverted mass ordering --- though it is disfavored --- significantly worsens the limits derived following our procedure.  This is because the standard-mixing region centered around $(\frac{1}{3}:\frac{1}{3}:\frac{1}{3})_\oplus$ lies very close to the present $1\sigma$ IceCube contour.  Thus, while under normal ordering the standard region lies outside the contour, under inverted ordering it is almost fully contained by it.  As a result, due to the hard $1\sigma$ cut implemented in our limit-setting procedure, changing the mass ordering has a large effect on the limits.  In contrast, limits derived using future flavor results, centered on $(\frac{1}{3}:\frac{1}{3}:\frac{1}{3})_\oplus$, would be marginally affected by the choice of mass ordering.

\setcounter{figure}{0}
\renewcommand{\thefigure}{D\arabic{figure}}
\begin{figure*}[t!]
 \centering
 \includegraphics[width=\columnwidth]{flavor_ratios_earth_var_v_em_enu_0100_tev_src_120_010_100_no_z_000_coded_by_v_1s_avg_nu_nubar.png}
 \includegraphics[width=\columnwidth]{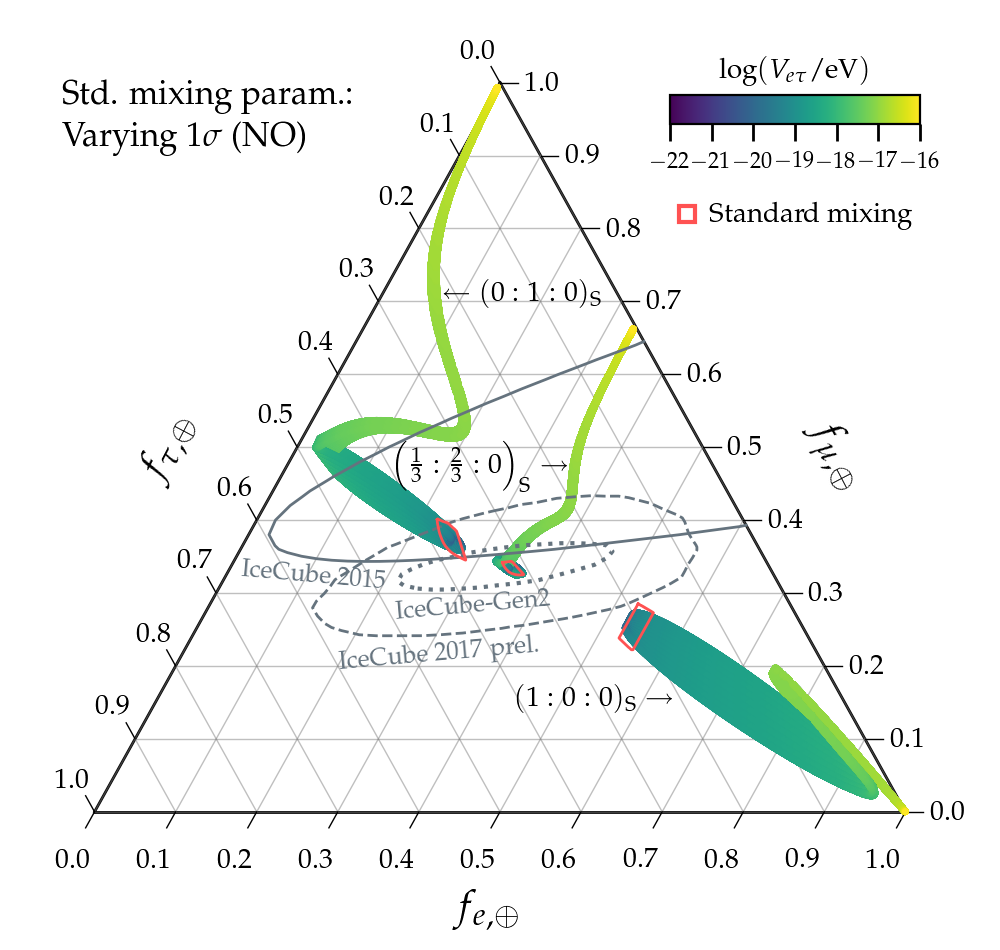}
 \includegraphics[width=\columnwidth]{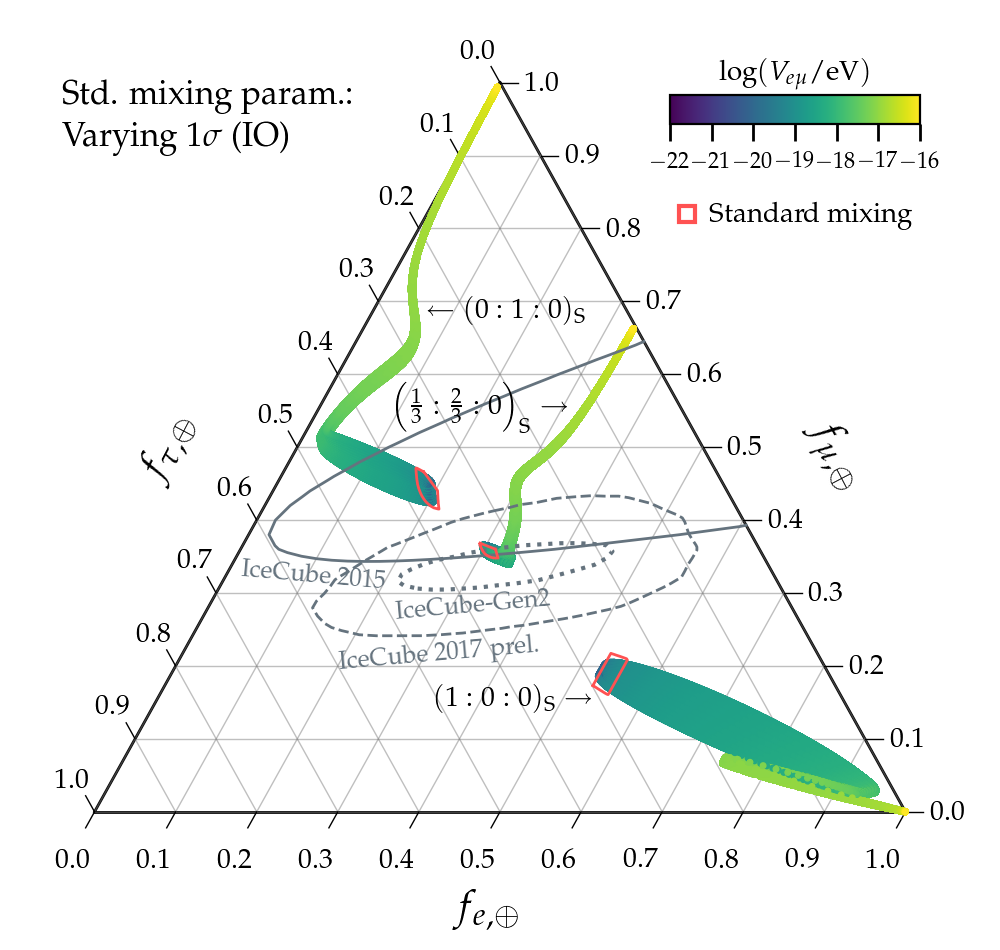}
 \includegraphics[width=\columnwidth]{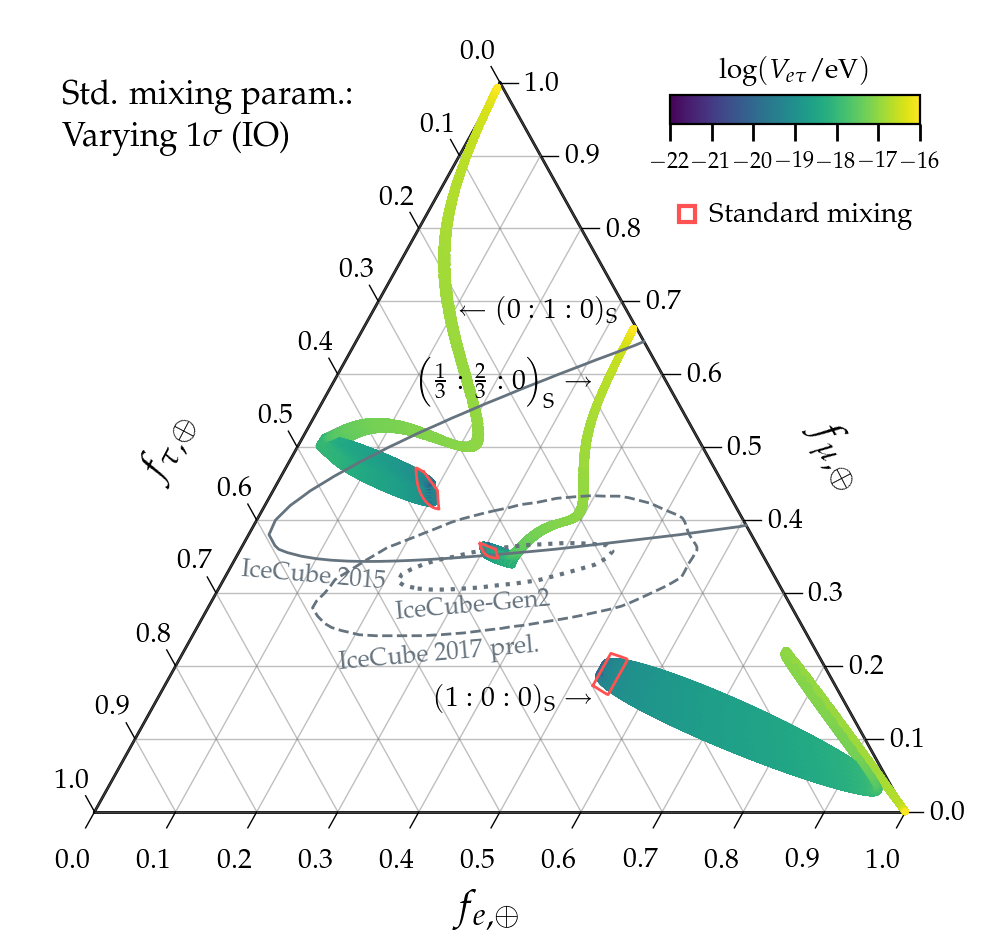}
 \caption{\label{fig:ternary_ordering_compare}Same as \figu{f_E_vary_V}, but for all possibilities of lepton-number symmetry and neutrino mass ordering: $L_e-L_\mu$ with normal ordering (NO, top left; same as \figu{f_E_vary_V}), $L_e-L_\tau$ with NO (top right), $L_e-L_\mu$ with inverted ordering (IO, bottom left), and $L_e-L_\tau$ with IO (bottom right).  Like in \figu{f_E_vary_V}, in these plots we fixed $E_\nu = 100$~TeV for illustration, but our limits are obtained using energy-averaged flavor ratios $\langle f_{\alpha,\oplus} \rangle$ (see main text), which behave similarly with $V_{e\beta}$.}
\end{figure*}

\begin{figure*}[b!]
 \centering
 \includegraphics[width=\columnwidth]{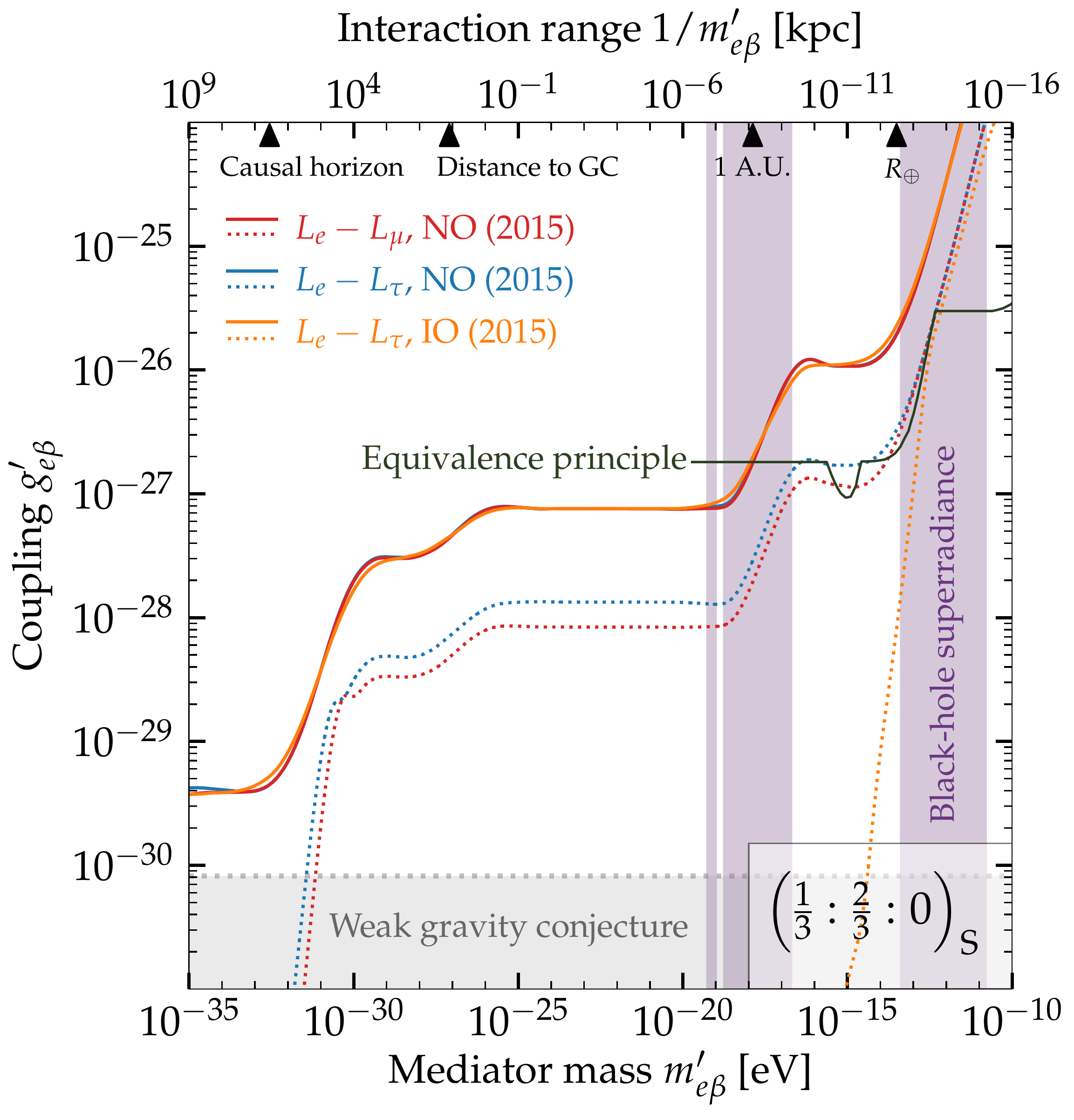}
 \includegraphics[width=\columnwidth]{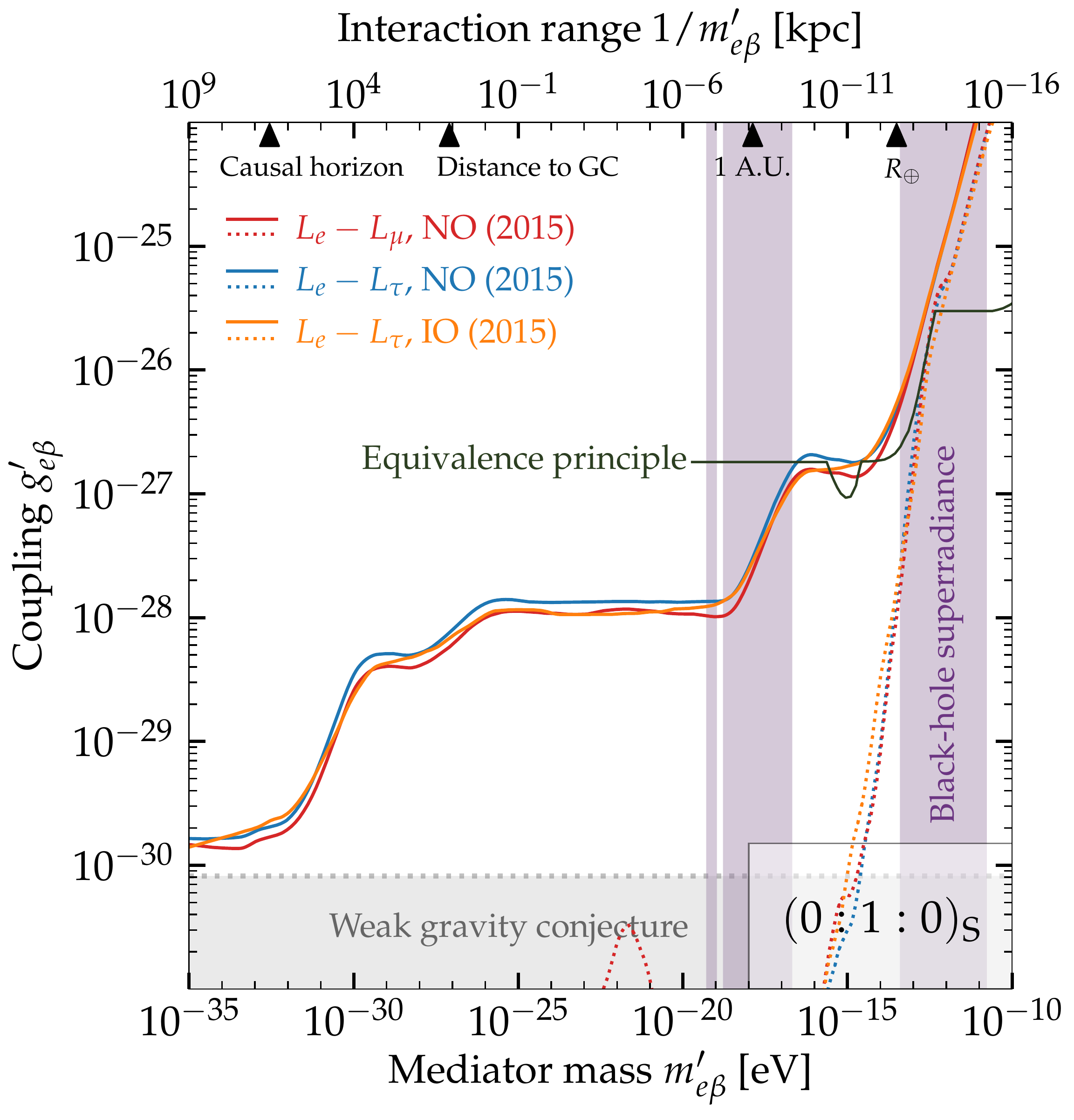}
 \caption{\label{fig:limits_ordering_compare}Constraints at $1\sigma$ on the mass and coupling of the $Z_{e\mu}^\prime$ and $Z_{e\tau}^\prime$ bosons, derived from current IceCube flavor measurements.  Like in \figu{limits_emu}, we assumed an astrophysical neutrino spectrum $\propto E_\nu^{-2.5}$. {\it Left:} Assuming the nominal expectation of flavor ratios $\left( \frac{1}{3}:\frac{2}{3}:0 \right)_{\rm S}$ at the source.  Here, upper-limit curves of $L_e-L_\mu$ and $L_e-L_\tau$ at NO are on top of each other.  {\it Right:} Assuming the alternative muon-damped ratios $(0:1:0)_{\rm S}$.}
\end{figure*}

\end{document}